\documentclass[twocolumn,twoside,slac_two]{revtex4}
\usepackage{graphicx}
\usepackage{fancyhdr}
\begin{document}

\title{Analysis of the Spectral Energy Distributions of $Fermi$ bright blazars}

%

\author{S. Cutini}
\affiliation{ASI Science Data Center, ASDC, ESRIN, Frascati, Italy}
\author{D. Gasparrini}
\affiliation{ASI Science Data Center, ASDC, ESRIN, Frascati, Italy}

\author{P. Giommi}
\affiliation{ASI Science Data Center, ASDC, ESRIN, Frascati, Italy}

\author{M.N. Mazziotta}
\affiliation{Istituto Nazionale di Fisica Nucleare, Sezione di Bari, 70126 Bari, Italy}

\author{C. Monte}
\affiliation{Istituto Nazionale di Fisica Nucleare, Sezione di Bari, 70126 Bari, Italy}
\affiliation{Dipartimento di Fisica "M. Merlin" dell'Universit\'a e del Politecnico di Bari, Italy}
\author{on behalf of Fermi LAT collaboration}

\begin{abstract}
 Blazars are a small fraction of all extragalactic sources but, unlike other objects, they are strong emitters across the entire electromagnetic spectrum. In this study we have conducted a detailed investigation of the broad-band spectral properties of the  gamma-ray selected blazars of the $Fermi$ LAT Bright AGN Sample (LBAS). By combining the accurately estimated $Fermi$ gamma-ray spectra with Swift , radio, NIR-Optical and hard-X/gamma-ray data, collected within three months of the LBAS data taking period, we were able to assemble high-quality and quasi-simultaneous Spectral Energy Distributions (SED) for 48 LBAS blazars.

\end{abstract}

\maketitle

\thispagestyle{fancy}


\section{Introduction}

The Large Area Telescope (LAT) on board of the Fermi Gamma Ray Space Telescope, launched  on 11 June 2008, provides unprecedented sensitivity in the $\gamma$-ray band (20 MeV to over 300 GeV, \cite{Atwood2009} with a large increase over its predecessors EGRET \cite{Thompson93}, and AGILE \cite{tavani08}. The  first three months of operations in sky-survey mode led to the compilation of a list of 205 $\gamma$-ray sources with statistical significance larger than 10$\sigma$ \cite{AbdoLATpaper}.

As largely expected from the previous results, most of the high Galactic latitude sources in this catalog are blazars \cite{AbdoAGNpaper}. 
%
One of the most effective ways of studying the physical properties of blazars is through the use of multi-frequency data. This approach has been followed assembling the Spectral Energy Distributions (SEDs)  of many radio, X-ray, and gamma-ray selected blazars. In all cases, however, the effectiveness of the  method was limited by the availability of only sparse, often non-simultaneous, flux measurements covering a limited portion of the electromagnetic spectrum. 
%

With Fermi, Swift, and other high-energy astrophysics satellites simultaneously on orbit, complemented by other space and ground-based observatories, it is now possible to assemble high-quality data to build simultaneous and well sampled SEDs of large and unbiased samples of AGN.

We derive the detailed SED of a subsample of 48 Fermi blazars using simultaneous or quasi-simultaneous data obtained from Swift and other ground and space based observatories.

\section{Multi frequency observations}

We describe the multi-frequency observations of LBAS blazars carried out between August and October 2008 with $Fermi$, and between May 2008 and January 2009 
with $Swift$ and other space and ground-based facilities.  

\subsection{Fermi LAT data analysis and gamma-ray energy spectra}

The Fermi-LAT data from 4 August to 31 October 2008 have been analyzed, selecting for each source only photons belonging to the 
diffuse class (Pass6 V3 IRF) \cite{Atwood2009}. Events within a $15^\circ$ Region of Interest (RoI) centered around the source have been selected. 
In order to discard photons from the Earth albedo, events with zenith angles larger than $105^\circ$ with respect to the Earth reference frame \cite{AbdoLATpaper} have been excluded from the data samples.

A maximum likelihood analysis (\textit{gtlike})\footnote{http://fermi.gsfc.nasa.gov/ssc/data/analysis/documentation/Cicerone} has been used to reconstruct the source energy spectrum.  
A  model is assumed for the source spectrum as well as for the diffuse background components, depending on a set of free parameters. The Galactic diffuse emission is modeled using GALPROP package 
 while the extragalactic  one is described by a simple power law \cite{AbdoLATpaper}.  The method has been implemented to estimate the parameters in each individual energy bin (2 bins per decade, starting  from 100 MeV), and the parameters obtained from the fits are used to  evaluate the sources fluxes. For each energy bin the source under investigation and all nearby sources in the RoI are described  by one parameter representing the integral flux in that energy  bin. The diffuse background components are modeled with one single parameter each, describing the normalization. For each bin, only fit results with a significance larger than $3\sigma$ have been retained. Depending on the flux and energy spectrum, 4 to 7 bins had positive detections for each AGN in the sample.

Once the differential flux in each energy bin $\phi(E)$ has been evaluated, the corresponding SED is then obtained by multiplying the differential flux by the square 
of the central energy value of that bin, i.e., $\nu F(\nu) = E^2 \phi(E)$ where $E=h \nu$. The vertical error bars 
represent only the statistical errors. The systematic uncertainties in the effective area for the Pass6 V3 DIFFUSE event selection have been estimated to be $10\%$ at 100 MeV, $5\%$ at 562 MeV and $20\%$ for energies greater than 10 GeV \citep{lsi61}.

\subsection{$Swift$ data analysis}

The $Swift$  Gamma-Ray-Burst (GRB) Explorer \citep{Gehrels04} is a multi-frequency, rapid response space observatory that was launched on 20 November 2004. To fulfill its purposes $Swift$ carries three instruments on board: the Burst Alert Telescope \citep[BAT,][]{Barthelmy05} sensitive in the 15-150 keV band, the X-Ray Telescope \citep[XRT,][]{Burrows05} sensitive in the 0.3-10.0 keV band, and the UV and Optical Telescope (170-600 nm) \citep[UVOT,][]{Roming05}. 
The very wide spectral range covered by these three instruments is of crucial importance for blazar issues as it covers where the transition between the 
synchrotron and inverse Compton emission usually occurs.

The primary objective of the $Swift$ scientific program is the discovery and rapid follow up of GRBs. However, there are
periods when $Swift$  is not engaged with GRB observations and the observatory 
can be used for different scientific purposes. The sources observed through this secondary science program are usually called $Swift$ fill-in targets. 
Since the beginning of its activities $Swift$ has observed hundreds of blazars as part of the fill-in program \citep[e.g.,][] {GiommiWMAP07}.
With the launch of AGILE and  $Fermi$, the rate of $Swift$  blazar observations increased significantly, leading to the observation (and detection) of all but  6 blazars in the LBAS sample.

The $Swift$ database currently includes 119 observations of 48 LBAS blazars that were carried out either simultaneously or within three months of the $Fermi$ LBAS data taking period. We used the UVOT, XRT and BAT data of these observations to build our SEDs. Some blazars were observed several times in the period that we consider in this paper; in such cases we considered only the exposures where the source was detected at minimum and maximum intensity by the XRT instrument.   

\subsubsection{UVOT data analysis}

$Swift$ observations are normally carried out so that UVOT produces a series of
images in each of the lenticular filters (V, B, U, UVW1, UVM2, and UVW2). 

Counts were
extracted from an aperture of 5'', radius for all filters and converted to
fluxes using the standard zero points \cite{poole08}. 

The fluxes were then de-reddened using the appropriate values of
$E(B-V)$ for each source taken from \cite{schlegel1998} with $A_{\lambda}/E(B-V)$
ratios calculated for UVOT filters using the mean interstellar extinction curve
from \cite{Fitzpatrick1999}. 

\subsubsection{XRT data analysis}

The XRT is usually operated in Auto State mode which automatically adjusts the
readout mode of the CCD detector to the source brightness, in an attempt to
avoid  pile-up (for details of the XRT observing modes)\cite{Burrows05}. 
Given the low count rate of our blazars most of the data were
collected using the most sensitive Photon Counting (PC) mode while Windowed Timing
(WT) mode was used for bright sources with shorter exposures.

The XRT data were processed with the XRTDAS software package (v.~2.4.1) developed 
at the ASI Science Data Center (ASDC) and distributed by the NASA High Energy 
Astrophysics Archive Research Center (HEASARC) within the HEASoft package (v.~6.6.1). 
Event files were calibrated and cleaned with standard filtering criteria with the 
{\it xrtpipeline} task using the latest calibration files available in the $Swift$ CALDB. 
Events in the energy range 0.3--10 keV with grades 0--12 
(PC mode) and 0--2 (WT mode) were used for the analysis.

Events for the spectral analysis were selected within a circle of 20 pixel 
($\sim$47 arcsec) radius, which encloses about 90\% of the PSF at 1.5 
keV \citep{Moretti05}, centered on the source position.
For PC mode data, when the source count rate is above $\sim$ 0.5 counts s$^{-1}$ data are 
significantly affected by pile-up in the inner part of the Point Spread Function (PSF). 
For such cases, after comparing the observed PSF profile with the analytical model derived by 
\cite{Moretti05}, we removed pile-up effects by excluding events detected within up to 6 pixels from 
the source position, and used an outer radius of 30 pixels. The value of the inner radius was evaluated individually 
for each observation affected by pile-up, depending on the observed count-rate. 

Source spectra were binned to ensure a minimum of 20 counts per bin to utilize the $\chi^{2}$ minimization 
fitting technique.

We fitted the spectra adopting an absorbed power law model with photon index $\Gamma_x$. When deviations 
from a single power law model were found, we adopted a log-parabolic law of the form 
$F(E)=K E^{(-a+b\cdot Log(E))}$ \cite{Massaro04} which has been shown to fit well the X-ray spectrum of 
blazars \cite[e.g.,][]{giommi05,Tramacere09}.  This spectral model is described by only two parameters: $a$, the photon index at 1 keV, and $b$, 
the curvature of the parabola. For both models the amount of hydrogen-equivalent column density (N$_H$) was fixed 
to the Galactic value along the line of sight \citep{Kalberla05}. 

\subsubsection{BAT hard X-ray data analysis}

We used survey data from the Burst Alert Telescope (BAT) on board $Swift$  to produce 15-200\,keV spectra of the blazars presented in 
this analysis. In order to do so, we used three years of survey data  \citep[see][for details]{ajello09} and extracted the spectra of those blazars
that are significantly detected in the 15--55\,keV band. Because of the very long integration time these data are not simultaneous with our $Fermi$ data.

Only 15 blazars, among those presented here, were detected by BAT at a significance $\geq 4$\,$\sigma$. The spectral extraction is performed as described in \cite{ajello08}  and the background-subtracted spectra represent the average emissions of the sources within the time spanned by the BAT survey.

\subsubsection{$Swift$ observations of LBAS blazars carried out before May 2008 or after January 2009}

The $Swift$ database includes a number of observations of LBAS blazars that were carried out outside the period that we consider useful to build our quasi-simultaneous SEDs.
These measurements are particularly important for the case of blazars that have never previously been observed by any X-ray astronomy satellite and were below the detection threshold of the ROSAT all sky survey.  When these $Swift$ observations have been analyzed and published by other authors we use the flux intensities reported in the literature, with particular reference to the latest on-line version of the  BZcat catalog \footnote{http://www.asdc.asi.it/bzcat}.  
For the cases where the $Swift$ results have not yet appeared in the literature we estimates the X-ray fluxes from the standard pipeline processing that is run at ASDC on all $Swift$ XRT data shortly after they are added to the archive. 

%

   
\subsection{Other multi-frequency data}

In order to improve the quality of our SEDs we complemented the $Fermi$ and $Swift$ quasi-simultaneous data with other multi-frequency flux measurements obtained 
from a number of on-going programs from ground and space-based observatories.
\subsubsection{Effelsberg radio observations}
 Quasi-simultaneous radio data for 25 sources of the first $Fermi$ bright source
catalog were obtained within a $Fermi$-related monthly broad-band monitoring
program including the Effelsberg 100-m radio telescope of the MPIfR
\cite[F-GAMMA project][]{Fuhrmann07,Angelakis08}. From this program, radio
spectra covering the frequency range 2.6 to 42\,GHz were selected to be within the
time period 4 August 2008 to 31 October 31 2008, i.e., quasi simultaneous to the $Fermi$ and $Swift$
observations.
\subsubsection{OVRO radio data}
 Quasi-simultaneous 15~GHz observations of 24 \emph{Fermi} LBAS sources
were made using the Owens Valley Radio Observatory (OVRO) 40~m
telescope.  These observations were made as part of an ongoing
$Fermi$-LAT blazar monitoring program.  In this program, all 1158
CGRaBS blazars north of declination $-20 degrees$ have been observed
approximately twice per week or more frequently since June~2007  
\cite{Healey08}. For each source, the maximum and minimum observed 15~GHz flux
densities during the 4~August to 31~October, 2008 period were included
in the quasi-simultaneous SEDs. 

\subsubsection{RATAN-600 1-22 GHz radio observations}

 Among the 48 objects for which we present $Swift$ and $Fermi$ simultaneous SEDs 32 were observed 
between September~10 and October 3, 2008 with the 600-meter ring radio telescope RATAN-600
\cite{RATANreview} of the Special Astrophysical Observatory, Russian
Academy of Sciences, located in Zelenchukskaya, Russia.
These observations, which produced  1--22~GHz instantaneous radio spectra, are part of a long-term program \cite[e.g.,][]{Kov02} to
monitor continuum spectra of active galactic nuclei with a strong
parsec-scale component of radio emission. 

\subsubsection{Radio, mm, NIR and optical data from the GASP-WEBT collaboration}
 In the period considered in this work, the GLAST-AGILE Support Program (GASP) originated from the Whole Earth Blazar Telescope (WEBT) \footnote{http://www/oato.inaf.it/blazars/webt/}   \cite[see e.g.,][]{villata07,raiteri08a})  carried out $\sim 3000$ optical ($R$ band) observations of 19 LBAS blazars, while $\sim 700$ near-IR ($JHK$, Campo Imperatore), and $\sim 600$ microwave (230 and 345 GHz, SMA) and radio data (5 to 43 GHz, Medicina, Noto, UMRAO) observations were taken on the same sources. 
In the SED plots we report the average, maximum and minimum values at each observed frequency in the period 4 August -- 31 October, 2008\footnote{Average flux densities were calculated on the 1-day binned data-sets, to avoid giving too much weight to the days with denser sampling.}.

\subsubsection{Mid-infrared VISIR observations}

 The MIR observations were carried out from 2006 to 2008 using VISIR
\cite{2004Lagage}, the ESO/VLT mid-infrared imager and spectrograph,
composed of an imager and a long-slit spectrometer covering several
filters in N and Q bands and mounted on Unit 3 of the VLT
(Melipal).  We performed broad-band photometry in 3 filters, PAH1
($\lambda$=8.59$\pm$0.42 $\mu$m), PAH2 ($\lambda$=11.25$\pm$0.59
$\mu$m), and Q2 ($\lambda$=18.72$\pm$0.88 $\mu$m) using the small
field in all bands (19.2'' x19.2'' and 0.075'' plate
scale). 

\subsubsection{Non-simultaneous Spitzer Space Telescope observations}
The Spitzer Space Telescope is a 0.85-meter class telescope launched on 25 August 2003.
 Spitzer Space Telescope  obtains images and spectra in the spectral range between 3 and 180 micron through three instruments on board: the InfraRed Array Camera (IRAC), which provides images at 3.6, 4.5, 5.8 and 8.0 microns, the Multiband Imaging Photometer for Spitzer (MIPS), which performs imaging photometry at 24, 70 and 160 micron, and the InfraRed Spectrograph (IRS) which provides spectra over 5-38 microns in low (R $\sim$ 60-127) and high (R $\sim $ 600) spectral resolution mode.
The Spitzer Science Archive include MIPS observations of 8 sources belonging to the LBAS sample, all of them performed earlier than three months from the start of LBAS 
data taking period. 
\subsubsection{AGILE $\gamma$-ray data}
 The $\gamma$-ray data collected by the Gamma-Ray Imaging
Detector (GRID) on board of AGILE  \cite{tavani08} for energies greater than 100 MeV used in this paper
(blue star symbols in the SED figures) are extracted from the First AGILE Catalog of high-confidence sources detected by 
AGILE during the first 12 months of operations,
from 9 July 2007 to 30 June 2008 \cite{Pittori09}. The differential AGILE flux values appearing in the SED figures at fixed energy point (E=300 MeV) have been rescaled from the mean flux 
above 100 MeV, obtained with a simple power law source model with fixed spectral index $-2.1$.

\section{Blazar SED observational parameters}
 We now estimate some key observational parameters that characterize the SED of our blazars, namely, the peak frequency and peak flux of the synchrotron component  ($\nu_p^S$ and $\nu_p^SF(\nu_p^S))$, and the peak frequency and flux of the inverse Compton part of the SED ($\nu_p^{IC}$ and $\nu_p^{IC} F(\nu_p^{IC}))$.

  
\begin{figure}
\includegraphics[height=7cm,angle=-90]{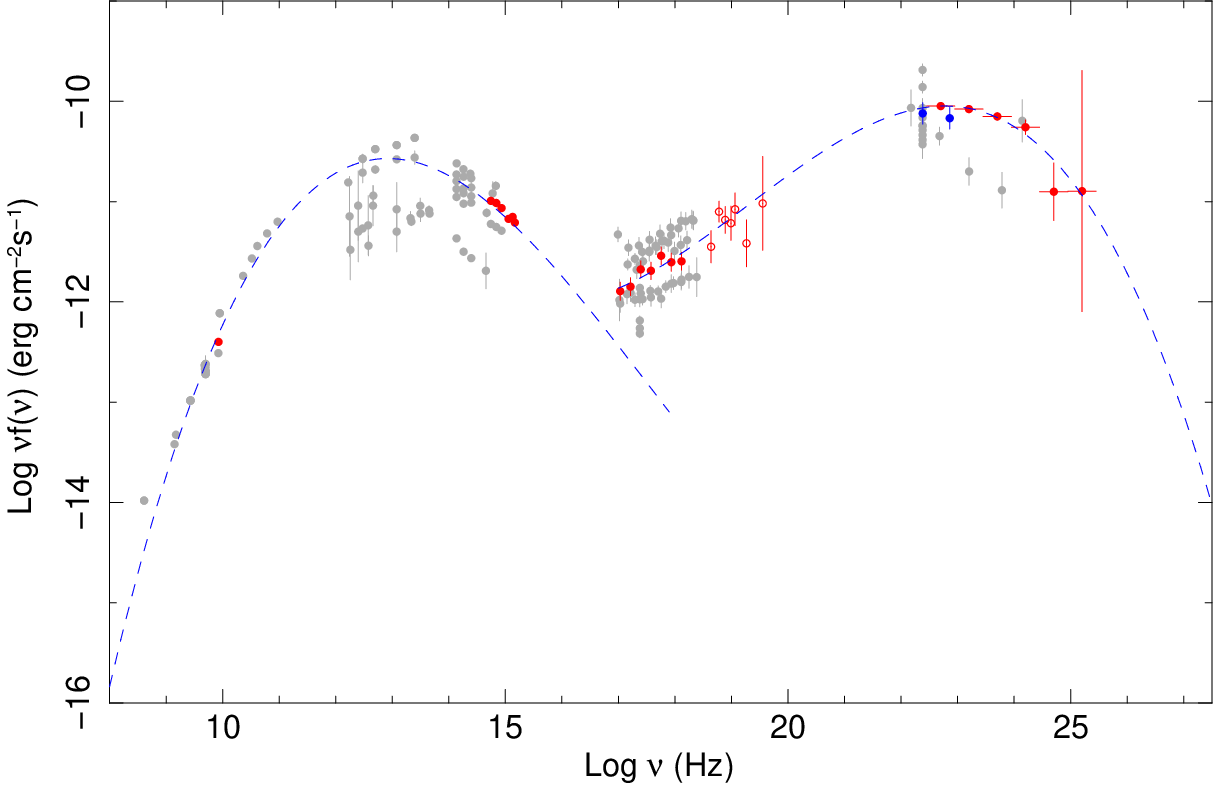}
\includegraphics[height=7cm,angle=-90]{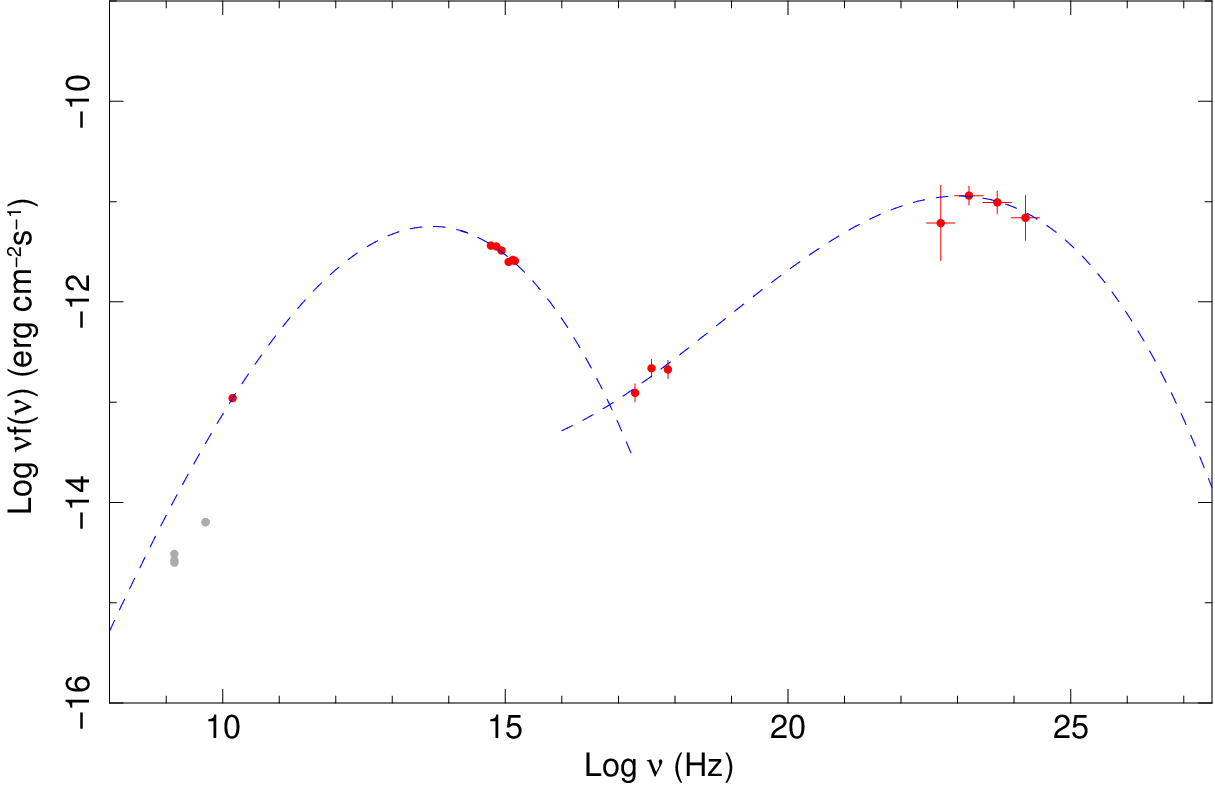}
\caption{The SED of  0FGL J0538.8-4403 = PKS0537-441. (left) and 
of 0FGL J0712.9+5034 = GB6 J0712+5033 (right)}
\label{fig:sed_aso0126}
\end{figure}

\begin{figure}
\includegraphics[height=7cm,angle=-90]{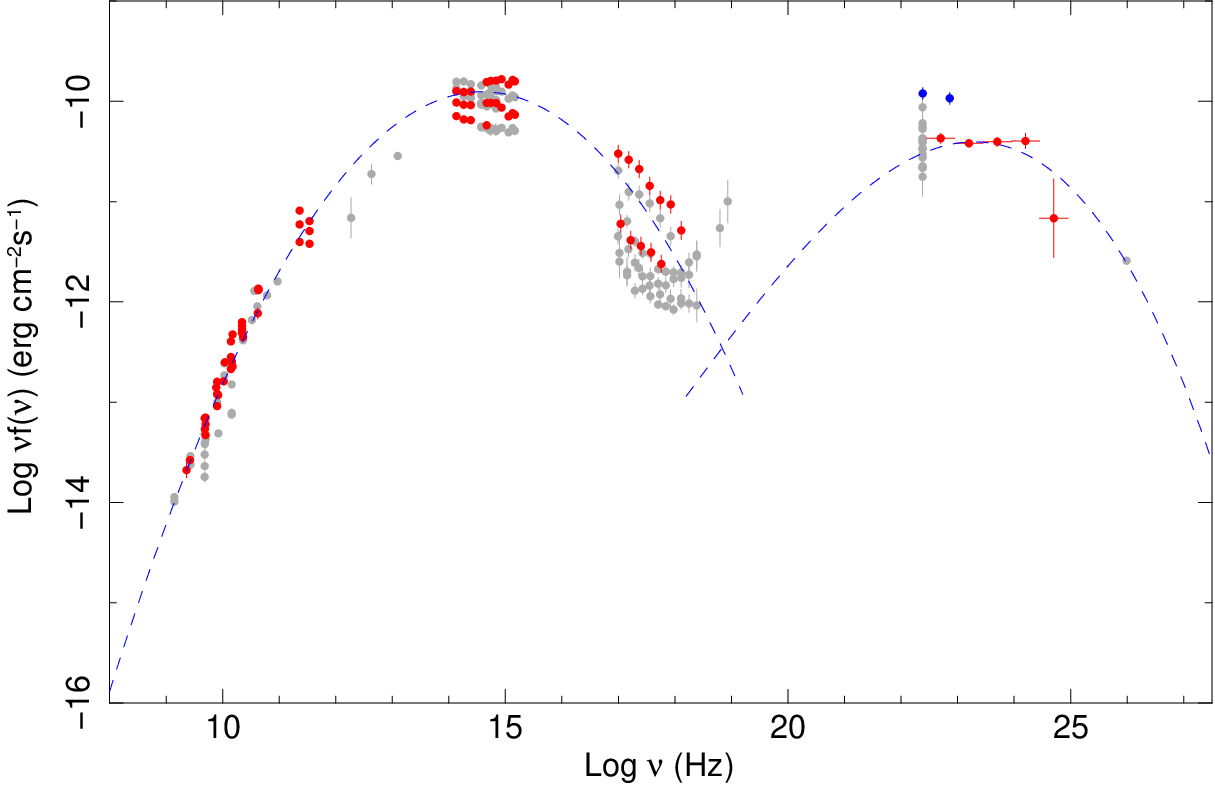}
\includegraphics[height=7cm,angle=-90]{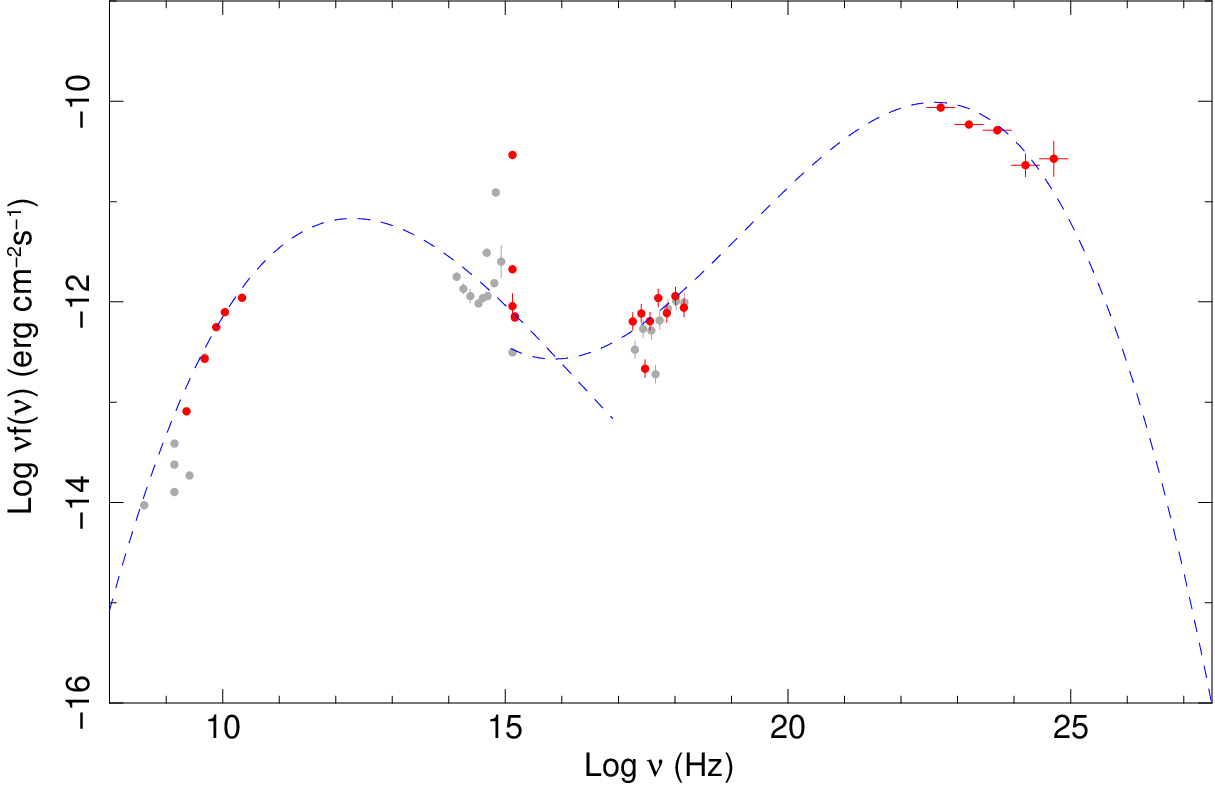}
\caption{The SED of 0FGL J0722.0+7120 =  S50716+714 (left) and 
of 0FGL J0730.4-1142 = PKS0727-11 (right)}
\label{fig:sed_aso0163}
\end{figure}

\begin{figure}
\includegraphics[height=7cm,angle=-90]{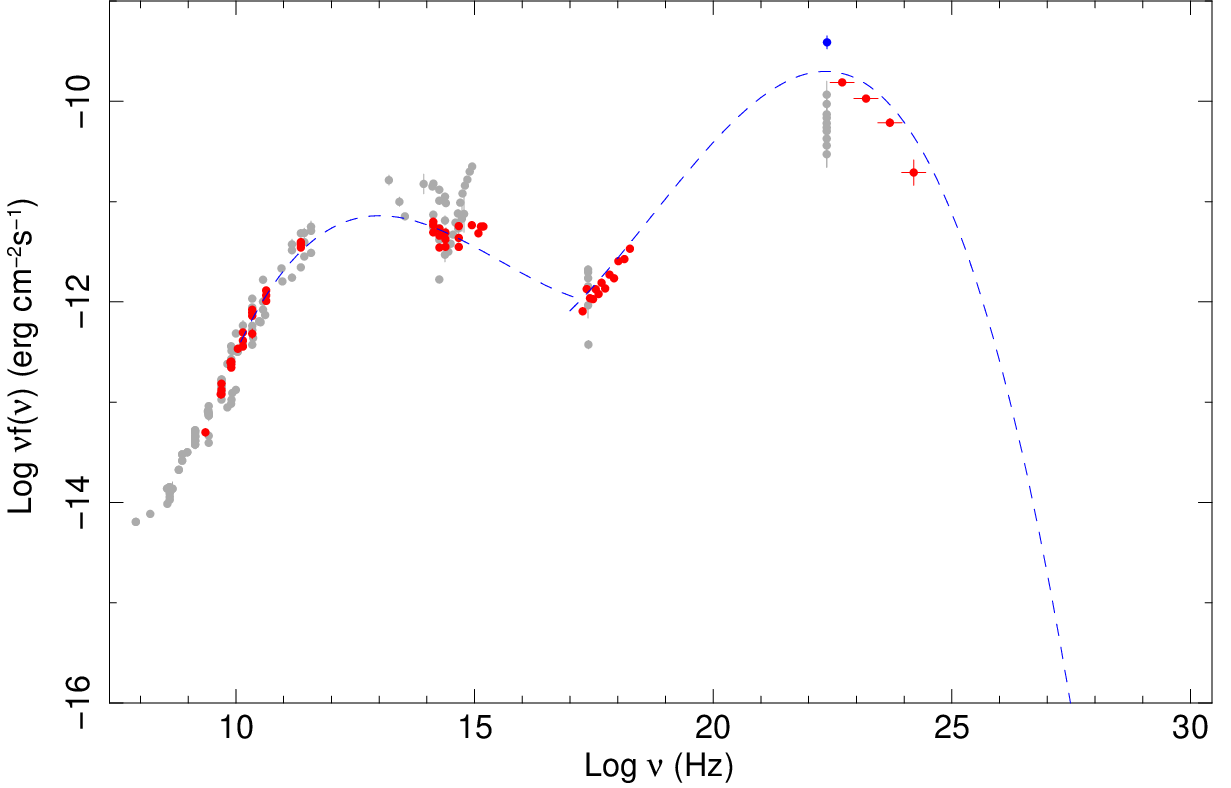}
\includegraphics[height=7cm,angle=-90]{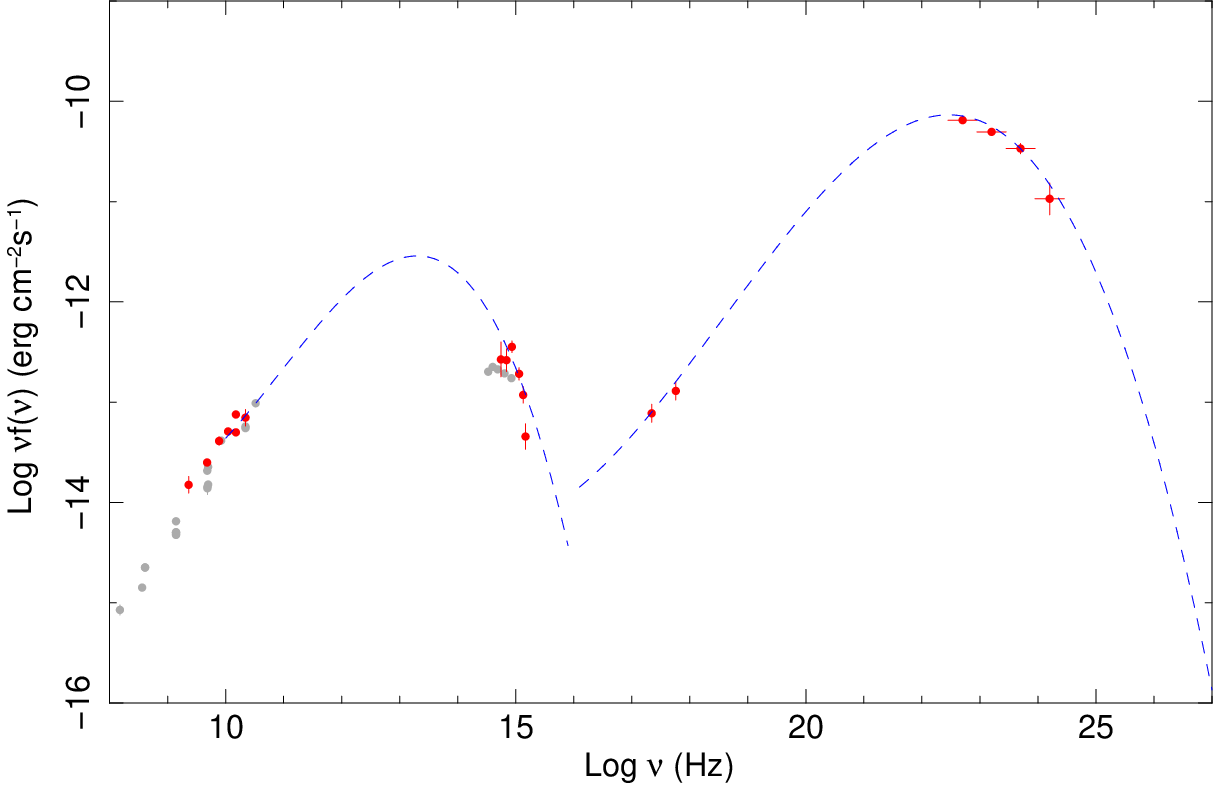}
\caption{The SED of 0FGL J1512.7-0905 =  PKS 1510-089 (left) and 
of 0FGL J1522.2+3143 = B2 1520+31 (right)}
\label{fig:sed_aso0326}
\end{figure}

\begin{figure}
\includegraphics[height=7cm,angle=-90]{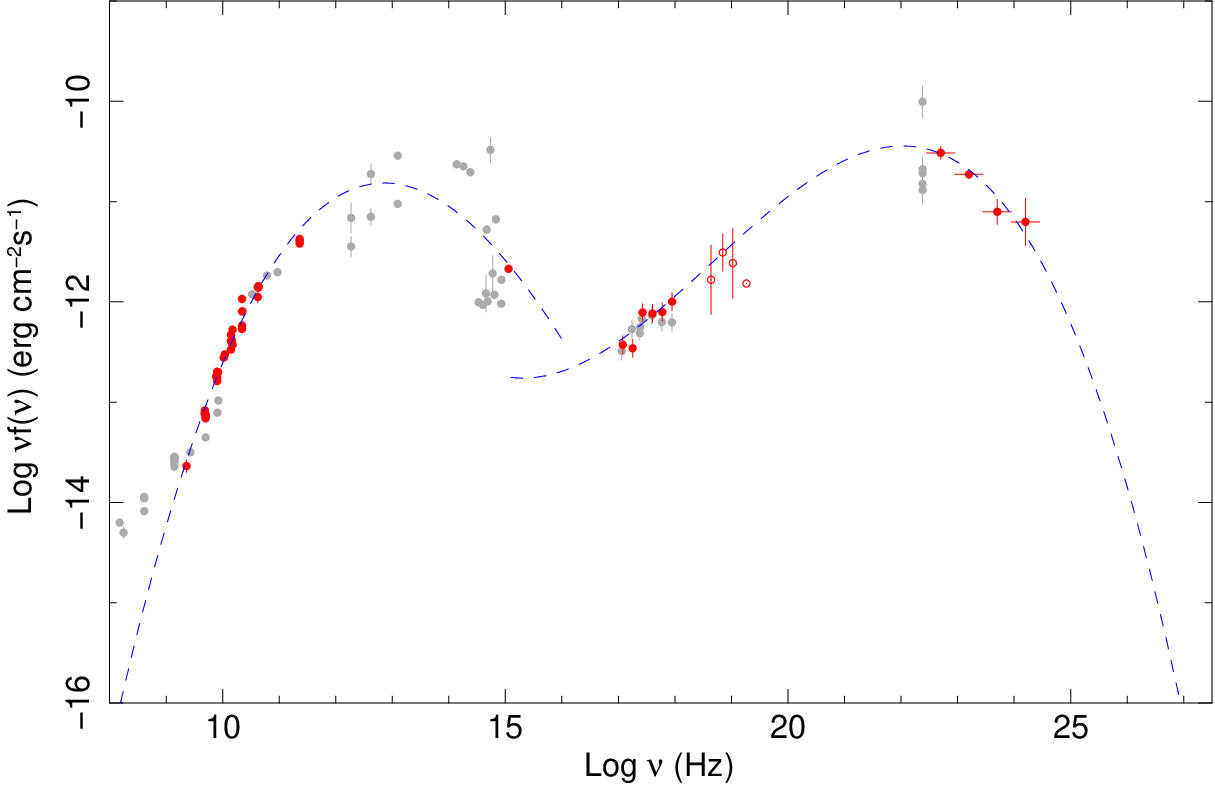}
\includegraphics[height=7cm,angle=-90]{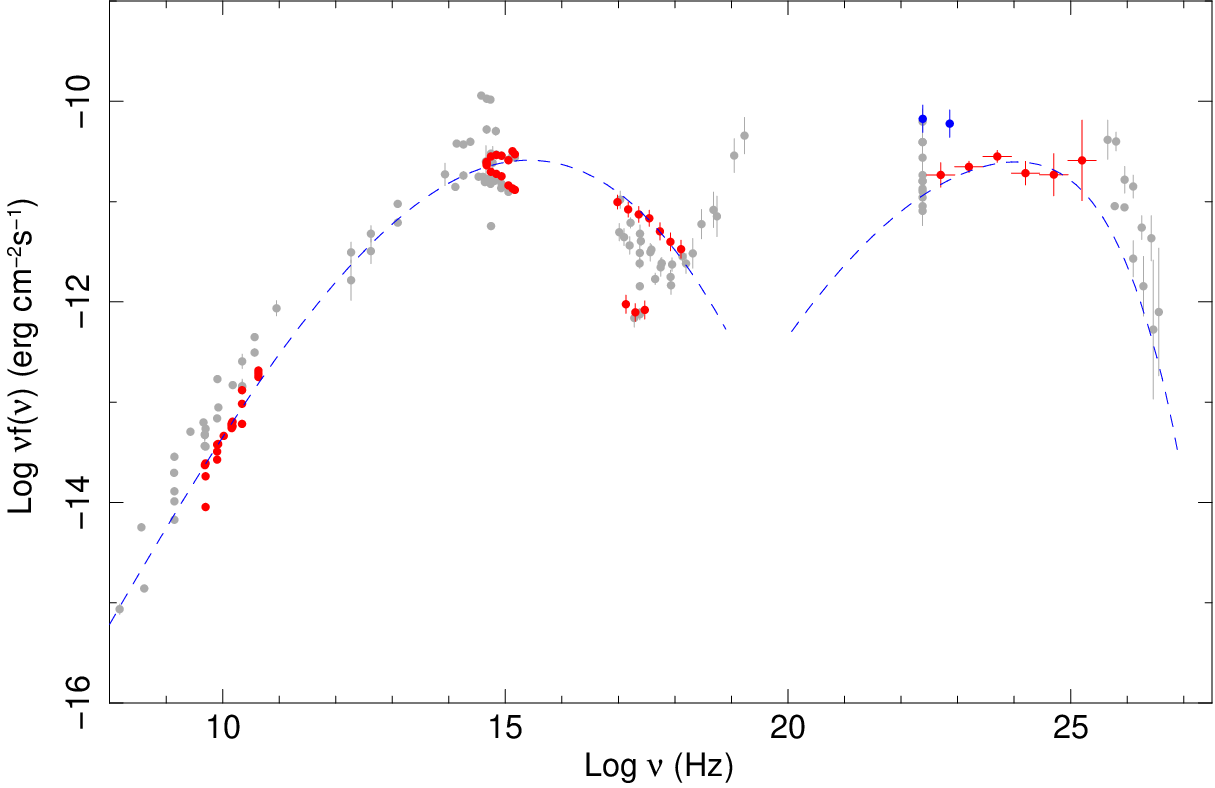}
\caption{The SED of 0FGL J1159.2+2912 = 4C29.45 (left) and 
of 0FGL J1221.7+2814 = ON231= W Comae (right)}
\label{fig:sed_aso0252}
\end{figure}

As fitting function we used a simple third degree polynomial:

\begin{equation}
\nu F_{\nu} = a\cdot \nu^3+b\cdot \nu^2+ c\cdot \nu+ d. 
\label{poly}
\end{equation}

In the case of high redshift sources (e.g.,  J1522.2+3143), we excluded from the fitting procedure all points in the optical/UV bands that are likely to be significantly affected by Lyman-$\alpha$ forest absorption.


We have estimated the peak of the inverse Compton power in the SED ($\nu_p^{IC}$) and the corresponding peak flux ($\nu_p^{IC} F(\nu_p^{IC}$)) by fitting the X-ray to $\gamma$-ray part of the SED,  which is dominated by inverse Compton emission using the polynomial function of Equation \ref{poly}.

There are some objects (e.g. J0722.0+7120  and  J1221.7+2814  ) in which the soft X-ray band is still dominated by synchrotron radiation, and only the $Fermi$ data can be used to constrain the inverse Compton component, so the above method is subject to large uncertainties. For this reason, in these cases, we have used the ASDC SED \footnote{http://tools.asdc.asi.it/SED/} interface to fit the simultaneous data points to a SSC model with a log-parabolic electron spectrum \cite{Tramacere09}. 

The best fit to both the synchrotron and inverse Compton components appear as dashed  lines in Figs.  \ref{fig:sed_aso0126}, \ref{fig:sed_aso0163}, \ref{fig:sed_aso0326} and \ref{fig:sed_aso0252}.
The details regarding the estimation of Spectral Energy Distribution observational parameters are accurately reported in the S. Rain\'o et al.``Interpretation of blazar SEDs based on broda-band quasi-simultaneous observations'' Proceedings.

\section{Conclusions}

We have carried out a detailed investigation of the broad-band (radio to high-energy) spectral properties of the LBAS sample of $Fermi$ bright blazars using a large number of 
multi-frequency simultaneous observations as well as literature and archival data.  Using data obtained with $Fermi, Swift$, radio/mm telescopes, infra-red, and optical facilities. 
For the first time high-quality quasi-simultaneous SED of blazars are available for a considerable number of blazars and this subset is representative of the entire LBAS.
We have been able to assemble simultaneous or quasi-simultaneous  SEDs of a sizable and representative fraction of an homogeneous sample ($\sim 45\%$) of blazars detected during an all sky survey .This collection of high-quality, well sampled, nearly simultaneous, broad-band SEDs for a large number of blazars is unprecedented and allowed us to estimate a number of important parameters characterizing the SED of selected blazars and to address some key aspects of blazar demographics and physics. Indeed we conclude that the distribution of the synchrotron peak frequency is very different for the FSRQ and BL Lac subsamples. In fact for FSRQs it starts at 
$\sim $10$^{12.5}$ Hz, peaks at  $\sim $10$^{13.3}$ Hz and it does not extend beyond $\approx$ 10$^{14.5}$ Hz, the distribution of BL Lacs is much flatter, starts at  $\sim $10$^{13}$ Hz and reaches much higher frequencies ( $\approx$ 10$^{17}$ Hz) than that of FSRQs.

\begin{acknowledgments}
The $Fermi$ LAT Collaboration acknowledges support from a number of agencies and institutes for both development and the operation of the LAT as well as scientific data analysis. These include NASA and DOE in the United States, CEA/Irfu and IN2P3/CNRS in France, ASI and INFN in Italy, MEXT, KEK, and JAXA in Japan, and the K.~A.~Wallenberg Foundation, the Swedish Research Council and the National Space Board in Sweden. Additional support from INAF in Italy and CNES in France for science analysis during the operations phase is also gratefully acknowledged.
This research is based also on observations with the 100 m telescope of the MPIfR (Max-Planck-Institut fur Radioastronomie) at Effelsberg. The
OVRO 40 m program is supported in part by NASA (NNX08AW31G) and the NSF (AST-0808050).
\end{acknowledgments}

\bigskip 

\begin{thebibliography}{9}   

\bibitem[Abdo et al. ~2009a]{AbdoLATpaper} Abdo, A.A. et al, 2009a, ApJS, 183,  46

\bibitem[Abdo et al. ~2009b]{AbdoAGNpaper} Abdo, A.A. et al, 2009b,  ApJ, 700, 597

\bibitem[Abdo et al. ~2009c]{lsi61} Abdo, A.A. et al., 2009c,  ApJ ~letters, submitted
\bibitem[Angelakis et al. ~2008]{Angelakis08} Angelakis, E., Fuhrmann, L., Marchili, N., Krichbaum, T.~P.,\& Zensus, J.~A. 2008, arXiv:0809.3912
\bibitem[Atwood et al. ~2009]{Atwood2009} Atwood, W.~B., et al. 2009, ApJ, 697, 1071
\bibitem[Ajello et al. ~2008]{ajello08} Ajello, M., \ et al.\ 2008, ApJ, 673, 96
\bibitem[Ajello et al. ~2009]{ajello09} Ajello, M., \ et al.\ 2009, arXiv:0905.0472 

\bibitem[Barthelmy et al. ~2005]{Barthelmy05} Barthelmy, S., Barbier, L.~M., Cummings, J., {et~al.} 2005, SSRv., 120, 95
\bibitem[Burrows et al. ~2005]{Burrows05} Burrows, D., Hill, J.~E., Nousek, J.~A., {et~al.} 2005, SSRv., 120, 165
\bibitem[Roming et al. ~2005]{Roming05} Roming, P. W.~A., Kennedy, T.~E., Mason, K.~O., {et~al.} 2005, SSRv, 120, 143
\bibitem[Fitzpatrick ~1999]{Fitzpatrick1999} Fitzpatrick, N. 1999 PASP, 111, 63
\bibitem[Fuhrmann et al. ~2007]{Fuhrmann07} Fuhrmann, L., Zensus, J.~A., Krichbaum, T.~P., Angelakis, E.,

\bibitem[Gehrels et al. ~2004]{Gehrels04} Gehrels, N. et al. 2004, ApJ, 611, 1005
\bibitem[Giommi et al.  ~2007]{GiommiWMAP07} Giommi, P., Capalbi, M.,  Cavazzuti, E.  et al. ~2007, A\&A, 468, 571



\bibitem[Healey et al. ~2008]{Healey08}Healey, S.\ E.\ et al.\ 2008, ApJ, 175, 97

\bibitem[Kalberla et al. ~2005]{Kalberla05}  Kalberla, P.M.W., Burton, W.B., Hartmann, Dap, et al. 2005, A\&A, 440, 775
\bibitem[Kovalev et al. ~2002]{Kov02} Kovalev, Y. Y., Kovalev, Yu. A., Nizhelsky, N. A., Bogdantsov, A. B.\ 2002, PASA, 19, 83
\bibitem[Korolkov \& Parijskij 1979]{RATANreview} Korolkov, D.~V., \& Parijskij, Yu.~N.\ 1979, Sky~Telesc., 57, 324
\bibitem[Lagage et al. ~2004]{2004Lagage} Lagage, P.O. et al., 2004., Msngr, 117, 12
\bibitem[Moretti et al. ~2005]{Moretti05} Moretti, A., Campana, S., Mineo, T., et al. 2005, Proceedings of SPIE, Vol. 5898, 360
\bibitem[Massaro et al.~2004]{Massaro04} Massaro, E., Perri, M., Giommi, P., \& Nesci, R. 2004, A\&A, 413, 489
\bibitem[Poole et al. ~2008]{poole08} Poole ? \ et al.\  2008 MNRAS, 383, 627
\bibitem[Raiteri et al. ~2008a]{raiteri08a} Raiteri, C.\ et al.\ 2008, A\&A 485, L17
\bibitem[Pittori et al. ~2009]{Pittori09} Pittori, C.\ et al.\  2009, A\&A, in press
\bibitem[Giommi et al.  ~2005]{giommi05} Giommi, P., Piranomonte, S.,  Perri, M. \& Padovani, P., 2005, A\&A, 434, 385 
\bibitem[Schlegel et al. 1998]{schlegel1998} Schlegel, D.J., Finkbeiner D.P., and Davis, M., 1998, ApJ, 500, 525
\bibitem[Tavani et al. 2008]{tavani08} Tavani, M., Barbiellini, G., Argan, A. et al. 2008, Nucl. Instr. and Meth. in Phys. Res. A, 588, 52
\bibitem[Tramacere 2009]{Tramacere09} Tramacere, A. et al., 2009, A\&A, 501, 879

\bibitem[Thompson et al. ~1993]{Thompson93}	Thompson, D. J. et al. ~1993 ApJSs, 86, 629




\bibitem[Villata et al. ~2007]{villata07} Villata M.\ et al.\  2007,  A\&A 464, L5


\end{thebibliography}

\end{document}